%% file: ms1.tex
\PassOptionsToPackage{hyphens}{url}
\documentclass{sig-alternate}
\usepackage{mathptmx} 

\usepackage{fancyhdr}
\usepackage[normalem]{ulem}
\usepackage[hyphens]{url}
\usepackage[sort,nocompress]{cite}
\usepackage[final]{microtype}
\usepackage[dvipsnames]{xcolor}
\usepackage[bookmarks=true,breaklinks=true,letterpaper=true,colorlinks,linkcolor=black,citecolor=blue,urlcolor=black]{hyperref}

\usepackage{booktabs} 
\usepackage{xspace}
\usepackage{enumitem}
\usepackage[binary-units=true]{siunitx}
\usepackage{amsmath}
\usepackage{ifthen}
\usepackage{graphicx}
\usepackage{multirow}
\usepackage{listings}
\usepackage{setspace}
\usepackage{float}
\usepackage[us,12hr]{datetime}
\usepackage{tikz}
\usepackage{anyfontsize}
\usepackage[keeplastbox]{flushend}

\fancypagestyle{firstpage}{
  \fancyhf{}
  
  \fancyhead[C]{} 
  \fancyfoot[C]{\thepage}
}

\newcommand{\paperlinespacing}[1]{0.7}

\newcommand{\change}[1]{{#1}}
\newcommand{\paratitle}[1]{\vspace{4pt}\noindent\textbf{#1.}}
\newcommand{\titleShort}{Polynesia\xspace}
\newcommand{\sgdel}[1]{}
\newcommand{\sgmod}[2]{#2}
\newcommand{\am}[1]{#1}
\newcommand{\sgii}[1]{{#1}}
\newcommand{\sgiii}[1]{{#1}}
\newcommand{\sgiv}[1]{{#1}}

\newcommand{\rev}[1]{{#1}}

\newcommand{\affilCMU}{$^\dag$}
\newcommand{\affilETH}{$^\diamond$}
\newcommand{\affilUIUC}{$^\ddag$}

\title{\vspace{-16pt}\titleShort: Enabling Effective Hybrid Transactional/Analytical Databases \\ with Specialized Hardware/Software Co-Design\vspace{-40pt}}
\author{%
\large{Amirali Boroumand\affilCMU\qquad%
Saugata Ghose\affilUIUC\qquad%
Geraldo F. Oliveira\affilETH\qquad%
Onur Mutlu\affilETH\affilCMU%
\vspace{5pt}}\\%
{\fontsize{11}{12}\selectfont\it%
\affilCMU Carnegie Mellon University\qquad%
\affilUIUC University of Illinois at Urbana--Champaign\qquad%
\affilETH ETH Z{\"u}rich%
}}

\begin{document}
\sloppy
\maketitle
\thispagestyle{firstpage}
\pagestyle{plain}

\setstretch{0.978}

\input{sections/abstract}
\input{sections/intro}
\input{sections/background}

\input{sections/motivation}

\input{sections/proposal}
\input{sections/update-propagation}

\input{sections/consistency}

\input{sections/analytic-engine}

\input{sections/methodology}
\input{sections/evaluation}

\input{sections/related}
\input{sections/conclusion}

{
\interlinepenalty=10000
\bibliographystyle{IEEEtranS}
\bibliography{references}
}

\end{document}

%% file: sections/abstract.tex

\begin{abstract}

\sgiii{An exponential growth in data volume, combined with increasing demand for 
real-time analysis (i.e., using the most recent data), has resulted in
the emergence of database systems that
concurrently support transactions and data analytics.
These \emph{hybrid transactional and analytical
 processing} (HTAP) database systems can support real-time data analysis
without the high costs of synchronizing across separate single-purpose databases.}
\sgdel{Unfortunately, given the high rate
at which data updates can occur in many domains, it is challenging
to perform real-time analysis without suffering a significant drop in
transactional and/or analytical throughput.}%
\sgii{Unfortunately, for many applications that perform a high
rate of data updates, state-of-the-art HTAP systems incur significant
drops in transactional (up to 74.6\%) and/or analytical (up to 49.8\%)
throughput compared to performing only transactions or only
analytics in isolation, due to
(1)~data movement between the CPU and memory,
(2)~data update propagation, and
(3)~consistency costs.}

We propose \emph{\titleShort}, 
a hardware--software co-designed \rev{system for in-memory HTAP databases}. 
\sgii{\titleShort
(1)~divides the HTAP system into \am{transactional and analytical processing islands,}
(2)~\sgiv{implements} custom algorithms and hardware to reduce the costs of update propagation and consistency, and
(3)~\am{exploits} processing-in-memory for the analytical islands to alleviate data movement.}
\sgii{Our evaluation shows that \titleShort outperforms three state-of-the-art
HTAP systems, with average transactional/analytical throughput improvements of
1.70X/3.74X, and reduces
energy consumption by 48\% over the prior lowest-energy system.}

\end{abstract}

%% file: sections/intro.tex

\section{Introduction}
\label{sec:intro}

Data analytics has become popular due to the exponential growth
in data generated annually~\cite{cisco-report}.
Many application domains have a critical need to perform 
\emph{real-time data analysis},
and make use of \emph{hybrid transactional and analytical processing}
(HTAP)~\cite{htap-gartner,sap-hana-evolution,htap}.
An HTAP database management system (DBMS) is a single-DBMS
solution that supports both transactional and analytical
workloads~\cite{htap-gartner,peloton,batchdb,htap-survey,real-time-analysis-sql}.
An \am{ideal} HTAP system should have three properties~\cite{batchdb}.
First, it should ensure that both transactional and analytical workloads 
\am{benefit from their own workload-specific optimizations (e.g., algorithms, data structures)}.
Second, it should guarantee data freshness \am{(i.e., access to \sgiv{the} most recent version of data)} for analytical workloads
while \am{ensuring \sgiv{that} both transactional and analytical workload have} a consistent view of data across the system.
Third, it should ensure that the latency and throughput of the transactional and
analytical workloads are the same as if they were run in isolation.

We extensively study state-of-the-art HTAP systems (Section~\ref{sec:bkgd:motiv}) 
and observe two key problems that prevent them from achieving all three
properties of an ideal HTAP system.
First, \rev{these} systems experience a drastic reduction in
transactional throughput (up to 74.6\%) and analytical throughput (up to 49.8\%) compared to
\am{when we run each in isolation.}
\am{This is because \sgiv{the mechanisms used} to provide data freshness and consistency}
induce a
significant amount of \emph{data movement} between the CPU cores and
main memory.
Second, HTAP systems often fail to provide effective performance isolation.
 These systems suffer from severe performance interference 
    because of the \sgii{high resource contention}
between transactional workloads and analytical workloads.
\emph{Our goal} in this work is to develop an HTAP system that
overcomes these problems while achieving all three of the
desired HTAP properties, \am{with new architectural techniques}.

\sgmod{Based on our observations, we}{We} propose a \rev{novel system 
for in-memory HTAP databases} called 
\titleShort.
\sgdel{To our knowledge, \titleShort is the first HTAP system to use
hardware--software co-design to enable real-time analysis and
meet all of the desired HTAP properties.}
\sgiv{The key insight behind \titleShort is to meet all three desired HTAP properties
by partitioning the computing resources into two isolated processing \emph{islands}:
\emph{transactional islands} and \emph{analytical islands}.}
Each island consists of
(1)~a replica of data for a specific workload, 
(2)~an optimized execution engine \am{(i.e., the \sgiv{software} that executes queries)}, and
(3)~a set of hardware resources (e.g., computation units, memory)
that cater to the execution engine and its memory access patterns.
\sgii{We co-design new software and specialized hardware support for the islands.}
This includes algorithms and accelerators for
update propagation (Section~\ref{sec:proposal:update-propagation}) and
data consistency (Section~\ref{sec:proposal:consistency}), and
a new analytical engine (Section~\ref{sec:proposal:analytic-engine})
that includes software to handle data placement
and runtime task scheduling, along with in-memory hardware for
task execution.

\sgii{In our evaluations \am{(Section~\ref{sec:eval})}, we show the benefits of each component of \titleShort,
and compare its end-to-end performance and energy usage to three
state-of-the-art HTAP systems.
\titleShort outperforms all three, with higher 
transactional (2.20X/1.15X/1.94X; mean of 1.70X) and 
analytical (3.78X/5.04X/2.76X; mean of 3.74X) throughput.
\titleShort consumes less energy than all three as well,
48\% lower than the prior lowest-energy system.
Overall, we conclude that \titleShort efficiently provides
high-throughput real-time analysis, by meeting all three
desired HTAP properties.}

%% file: sections/background.tex

\section{HTAP Background}
\label{sec:bkgd:requirements}

To enable real-time analysis, we need a DBMS that 
allows us to run analytics 
on fresh data ingested by transactional engines. 
\sgmod{Traditionally,
transactional and analytical workloads ran on separate specialized DBMSs, where
each DBMS' engine and data format was optimized specifically for the workload,
     and updates are periodically propagated from the 
transactional DBMS to the analytical DBMS, using a complex process called ETL.
The main drawback of this approach is low
data freshness, as the ETL is performed only on the order of hours or
days.
As a result, analytical workloads on a traditional system could not practically
perform real-time analysis
~\cite{peloton, batchdb, hyper, htap-gartner}.

  In recent years, several}{Several} works from industry (e.g., \cite{sap-hana, oracle-dual-format, sql-htap,sap-soe, real-time-analysis-sql}) 
 and academia (e.g., \cite{hyper,peloton,hyrise,h2tap,l-store,batchdb,scyper, janus,janus-graph,sap-parallel-replication}
  attempt to address \sgmod{this issue and propose}{issues with data freshness by proposing} various techniques to 
support both transactional workloads and analytical workloads in a \emph{single} database system. 
This combined approach is known as \emph{hybrid transactional and analytical processing} (HTAP).
\label{sec:bkgd:htap-requirements}
To enable real-time analysis, an HTAP system should exhibit several key properties~\cite{batchdb}:

\paratitle{Workload-Specific Optimizations}
The system should provide each workload with optimizations specific to
the workload. 
Each workload requires different algorithms and data structures,
based on the workload's memory access patterns,
 to achieve high throughput and performance. 

\paratitle{Data Freshness and Data Consistency} The system should always provide the analytics workload
 with the \emph{most recent version} of data, even when transactions keep updating the
 data at a high rate. Also, the system needs to guarantee data consistency
 across the entire system, such that analytic queries observe a consistent view of data,
regardless of the freshness of the data.

\paratitle{Performance Isolation} The system should ensure that the latency
 and throughput of either workload is not impacted by running them concurrently
 within the same system.

\vspace{3pt}
Meeting all of these properties at once is very challenging, as these two workloads have different
underlying algorithms and access patterns, and optimizing for one property can often require a trade-off
in another property.

%% file: sections/motivation.tex

\section{Motivation}
\label{sec:bkgd:motiv}

There are two major types of HTAP systems:
(1)~single-instance design systems and
(2)~multiple-instance design systems.
In this section, we study both types, and analyze why neither type 
can meet all of the desired properties of an HTAP system (as we
describe in Section~\ref{sec:bkgd:htap-requirements}).

\subsection{Single-Instance Design}
\label{sec:bkgd:single}

One way to design an HTAP system is to maintain a single instance of the data, 
which both analytics and transactions work on, to ensure that
analytical queries access the most recent version of data. 
Several HTAP proposals from academia and industry
are based on this approach~\cite{hyper, peloton, l-store, hyrise, h2tap, sap-hana}. 
While single-instance design enables high data freshness, we find that it
suffers from three major challenges:

\paratitle{(1)~High Cost of Consistency and Synchronization}
\sgmod{Since analytical and transactional workloads 
work on the same instance of data concurrently, single-instance-based systems 
need to ensure that the data is consistent and synchronized. 
One approach to consistency is to let both
transactions and analytics work on the same copy of data, and use locking
protocols~\cite{locking} to maintain consistency across the system. However,
locking
has two major drawbacks. First, it
significantly reduces the update throughput of transactions~\cite{virtual-snapshot-nosql,hyper2}
and degrades data freshness,
 as it blocks transactions from updating objects that
 are being read by long-running analytics queries. 
 Second, it can frequently
 blocks analytics, as with the high update rate of transactions,
 a transactional workload can often lock out the analytics workload,
leading to a significant drop in throughput.
To avoid these drawbacks,}{To avoid the throughput bottlenecks
incurred by locking protocols~\cite{locking},}
 single-instance-based HTAP systems resort to either 
 snapshotting~\cite{hyper,h2tap,scyper,ankerdb} or multi-version concurrency
 control (MVCC)~\cite{hyper2, peloton}.
Unfortunately, \sgmod{we find that snapshotting and MVCC}{both solutions} have significant
drawbacks of their own.

\emph{Snapshotting:} Several HTAP systems (e.g., \cite{hyper, scyper, h2tap}) 
use a variation of multiversion synchronization, called snapshotting, to
 provide consistency via snapshot isolation~\cite{isolation-level,serializable-isolation}.
 Snapshot Isolation guarantees that all reads in a
 transaction see a consistent snapshot of the database state, which is
 the last committed state before the transaction started. These systems explicitly
 create snapshots from the most recent version of operational data, and let the analytics
 run on the snapshot while transactions continue updating the data. 

\begin{figure}[h]
    \vspace{-10pt}
    \centering
        \centering
        \includegraphics[width=\linewidth]{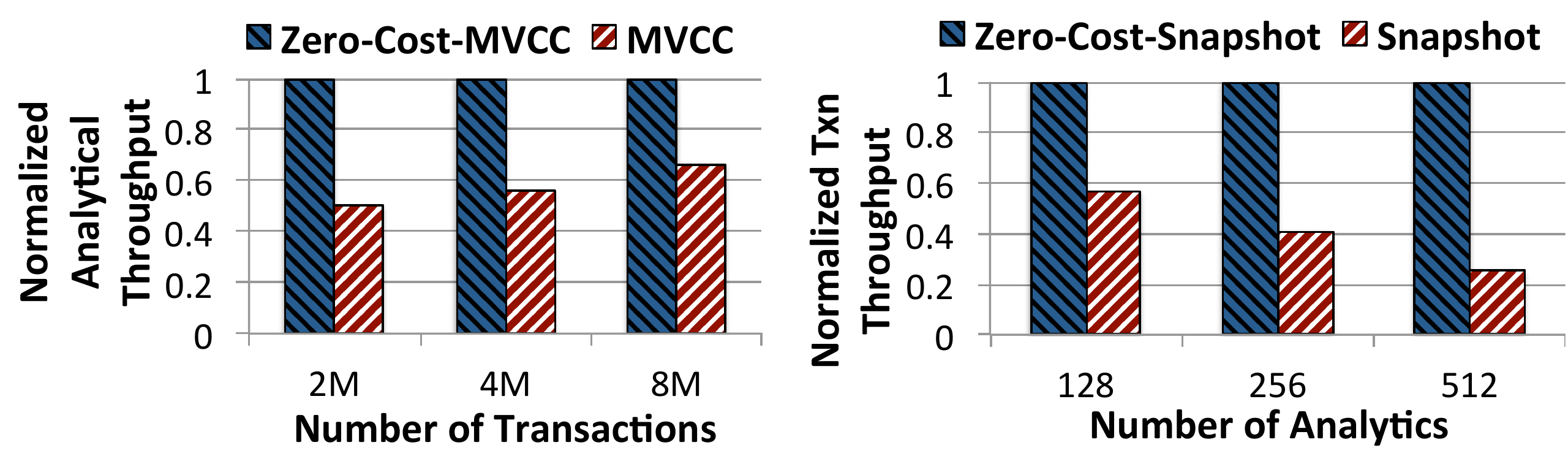}%
\vspace{-22pt}
    \caption{Effect of MVCC on analytical throughput (left) and snapshotting on transactional throughput (right).}
    \label{fig:motivation-consistency}
    \vspace{-5pt}
\end{figure}

 We analyze the effect of state-of-the-art snapshotting~\cite{software-snapshotting,h2tap} 
 on throughput, for an HTAP system with two transactional and two analytical
 threads (each runs on a separate CPU). 
Figure~\ref{fig:motivation-consistency} (right)
shows the transaction throughput with snapshotting, normalized to a
zero-cost snapshot mechanism, 
for three different rates of analytical queries.
We make two observations from the figure.
First, at 128~analytical queries, snapshotting reduces throughput by 43.4\%.
Second, the throughput drops as more analytical queries are being performed, with a drop of
74.6\% for 512~analytical queries.
We find that the majority of this throughput reduction occurs because \texttt{memcpy} is
used to create each snapshot, which introduces significant interference among the
workloads and generates a large amount of data movement between the CPU and main
memory.
The resulting high contention for shared hardware resources  (e.g., off-chip channel, 
memory system) directly hurts the throughput.

\emph{MVCC:} 
While both MVCC and snapshotting provide snapshot isolation, 
MVCC avoids making full copies of data for its snapshot.
In MVCC, \sgdel{instead of replacing the old data on an update,}
the system keeps several versions of the data in each entry.
The versions are chained together with pointers, and contain timestamps.
Now, instead of reading a separate snapshot, an analytics query can
simply use the timestamp to read the correct version of the data,
while a transaction can add another entry to the end of the chain
without interrupting.
As a result, updates never
 block reads, which is the main reason that MVCC has
 been adopted by many transactional DBMSs (e.g., \cite{sap-hana,oracle-dual-format,sql-htap}).

However, MVCC is not a good fit for mixed analytical and transactional workloads in HTAP.
We study the effect of MVCC 
on system throughput, using
the same hardware configuration that we used for snapshotting. 
 Figure~\ref{fig:motivation-consistency} (left) shows 
the analytical throughput of MVCC, normalized to a zero-cost version of MVCC,
for a varying transactional query count.
 We observe that the analytical throughput significantly decreases (by 42.4\%) compared to zero-cost MVCC. 
 We find that long version chains are the root cause of the throughput reduction.
 Each chain is organized as a linked list, 
and the chains grow long as many transactional queries take place. 
 Upon accessing a data tuple, the analytics query traverses a lengthy version 
 chain, checking each chain entry's timestamp to locate
 the most recent version that is visible to the query. As analytic queries touch a large
 number of tuples, this generates a very large number of random memory accesses,
leading to the significant throughput drop.

\paratitle{(2)~Limited Workload-Specific Optimization} 
A single-instance design severely limits workload-specific optimizations,
\sgmod{as it is very challenging to enable optimizations for both types of workloads on 
 a single replica of data.}{as the instance cannot have different optimizations for each workload.}
 Let us examine the data layout in relational databases as an example.
 Relational transactional 
 engines use a row-wise or N-ary storage model (NSM) for data layout,
 as it provides low latency and high throughput for update-intensive queries~\cite{kallman2008h}.
 Relational analytics engines, on the other hand, employ a column-wise or Decomposition Storage Model (DSM)
  to store data, as it provides better support for
 columnar accesses, compression, and vectorized execution. 

It is inherently impossible for a single-instance-based system to implement both 
formats simultaneously, and many such systems simply choose one of the 
layouts~\cite{hyper, l-store, h2tap}.
 A few HTAP systems attempt to provide a
 hybrid data layout (e.g., \cite{peloton}) or multiple data layouts in
 a single replica (e.g., \cite{oracle-dual-format}). However, these
 systems need to periodically convert data between different data
 formats, which leads to significant overhead and compromises
 data freshness~\cite{htap-survey}.

\paratitle{(3)~Limited Performance Isolation}
It is critical to ensure that running
analytics queries alongside transactions does not violate strict transactional latency and
throughput service-level agreements. 
Unfortunately, running both on the same instance of data, and sharing hardware resources,
leads to severe performance interference. We evaluate the effect of performance
interference using the same system configuration that we used for snapshotting and MVCC. 
Each transactional thread executes
2M queries, and each analytical thread runs 1024 analytical queries. 
We assume that there is no cost for consistency and synchronization.
Compared to running transactional queries in isolation, the transactional throughput
drops by 31.3\% when the queries run alongside analytics.
 This is because analytics are very data-intensive and generate a large
 amount of data movement,
 which leads to significant 
 contention for shared resources (e.g., memory system, off-chip bandwidth).
Note that the problem worsens with realistic consistency mechanisms, 
 as they also generate a large amount of data movement.

\subsection{Multiple-Instance Design}
\label{sec:bkgd:multiple}

The other approach to design an HTAP system is to maintain multiple instances of the data using
 replication techniques, and dedicate and optimize each instance to a specific
 workload (e.g., \cite{batchdb,oracle-dual-format,sql-htap,sap-soe,scyper,janus,htap-survey}). 
 Unfortunately, multiple-instance \sgdel{design }systems
suffer from \sgmod{a number of}{several} challenges:

\paratitle{Data Freshness} One of the major challenges in multiple-instance-based approach 
is to keep analytical replicas up-to-date even when the transaction update rate is high, 
without compromising performance isolation~\cite{htap,batchdb}. To maintain data freshness, the
system needs to (1)~gather updates from transactions and ship them to
analytical replicas, and (2)~perform the necessary format conversion and apply the
updates. 

\begin{figure}[h]
    \vspace{-5pt}
    \centering
        \centering
        \includegraphics[width=\linewidth]{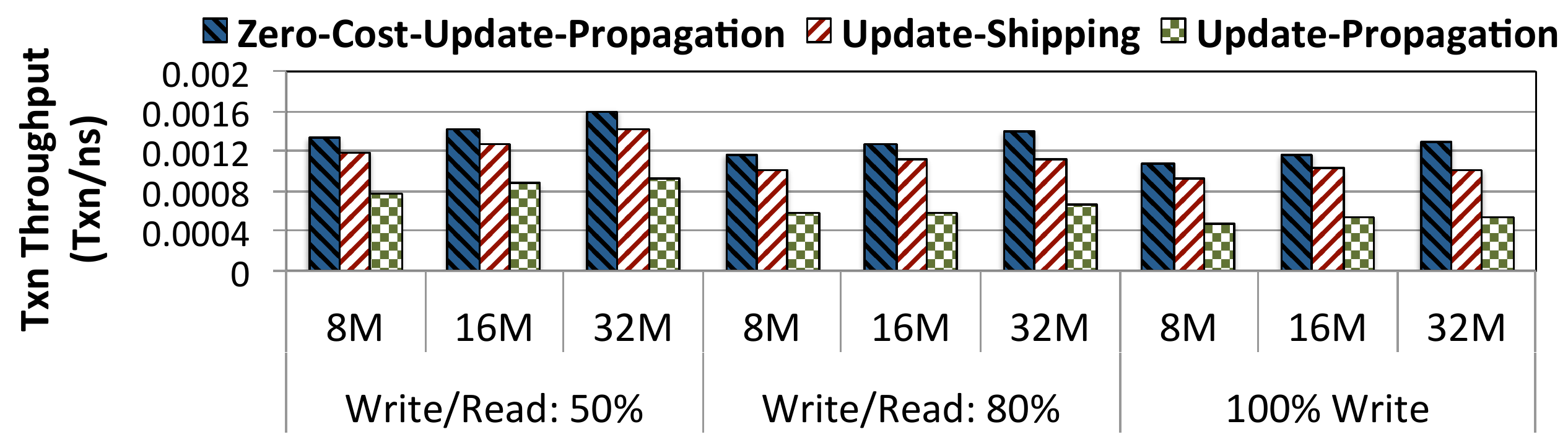}%
\vspace{-22pt}
    \caption{Transactional throughput across different number of transactions and different write intensities.}
    \label{fig:motivation-update-propagation}
    \vspace{-5pt}
\end{figure}

\begin{figure}[h]
    \vspace{-10pt}
    \centering
        \centering
        \includegraphics[width=\linewidth]{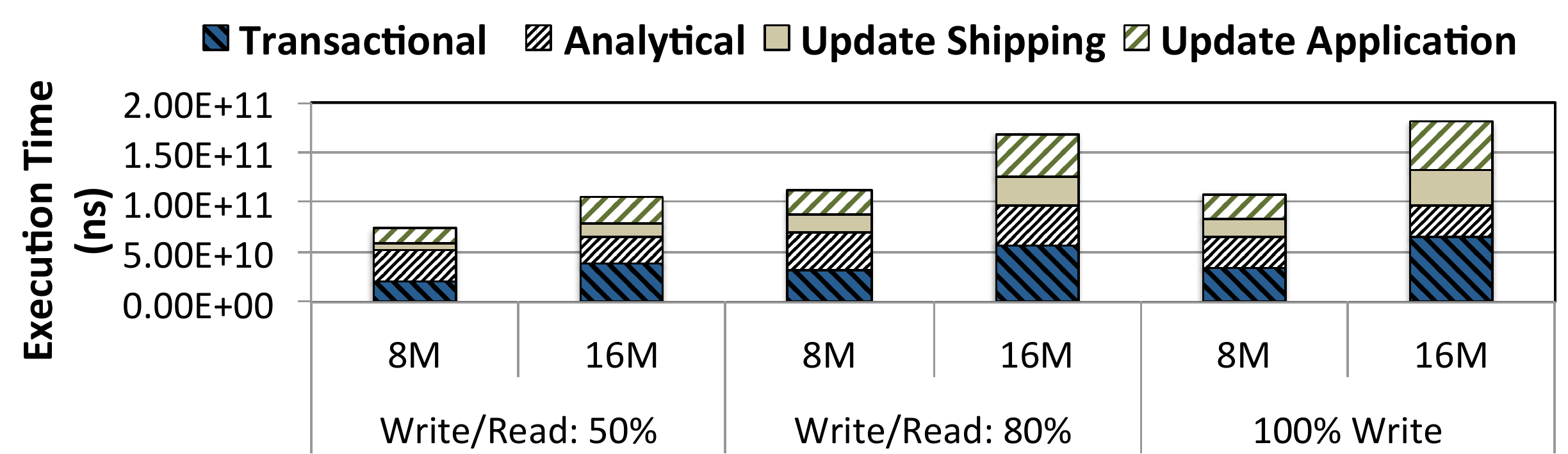}%
\vspace{-22pt}
    \caption{Execution time breakdown across different number of transactions and different write intensities.}
    \label{fig:motivation-update-propagation-latency}
    \vspace{-5pt}
\end{figure}

\emph{Gathering and Shipping Updates:} 
Given the high update rate of transactions,
 the frequency of the gathering and shipping process has a direct effect on data freshness. 
 During the update shipping process, the system needs to (1)~gather updates
 from different transactional threads, (2)~scan them to identify the target location
 corresponding to each update, and (3)~transfer each update to the corresponding location.
 We study the effect of update shipping on transactional throughput
 for a multiple-instance-based HTAP system 
 (see Section~\ref{sec:methodology}). 
 Our system has two transactional and two analytical threads (each running on a CPU core).
Figure~\ref{fig:motivation-update-propagation}
 shows the transactional throughput 
for three configurations: (1)~a baseline with zero
cost for update shipping and update application,
 (2)~a system that performs only update shipping, and (3)~a system that performs both update shipping 
 and update application (labeled as \emph{Update-Propagation}).
 We observe from the figure that the transactional throughput of update shipping reduces by 14.8\% 
compared to zero-cost update shipping and application.
 Our analysis shows that when the transactional queries are more update-intensive, the overhead becomes
 significantly higher. For update intensities of 80\% and 100\%, the throughput drops further
(by 19.9\% and 21.2\%, respectively). 
We find that \sgmod{the reduction in throughput}{this reduction} is mostly because the update shipping process generates a large amount of 
data movement and takes several CPU cycles. Figure~\ref{fig:motivation-update-propagation-latency} 
shows the breakdown of execution time during update propagation process. We find that update shipping 
accounts for on average 15.4\% of the total execution time.

\emph{Update Application}: 
The update application process can be very challenging, due to the need to transform
updates from one workload-specific format to another.
 For example, in relational analytics, the analytics engine uses several optimizations to speed up long-running scan queries
 and complex queries with multiple joins. 
  To minimize the amount of data that
 needs to be accessed, analytics engines employ DSM
 representation to store data~\cite{c-store},
and can compress tuples using an order-preserving
 dictionary-based compression (e.g., dictionary encoding~\cite{numa-scan,dict-compression,compression-c-store}).

 In our example, a single tuple update, stored in the NSM layout by the transactional workload,
requires multiple random accesses to apply the update 
 in the DSM layout. 
Compression further complicates this, as columns may need to be decompressed, updated, and
 recompressed. For compression algorithms that use sorted tuples, such as
dictionary encoding, the updates can lead to expensive shifting
 of tuples as well. 
 These operations generate a large amount of data movement and spend many CPU cycles. 
The challenges \sgdel{of update application }are significant enough that some prior works
give up on workload-specific optimization to try and maintain performance~\cite{batchdb}.

 From Figure~\ref{fig:motivation-update-propagation}, we observe
 that the update application process reduces the transactional throughput of the
 Update-Propagation configuration by 49.6\% \sgmod{compared to}{vs.} zero-cost update propagation.
 As the write intensity increases (from 50\% to 80\%), the throughput suffers
 more (with a 59.0\% drop at 80\%). This is because 23.8\% of the CPU cycles (and 30.8\% of cache misses)
 go to the update application process (Figure~\ref{fig:motivation-update-propagation-latency}), 
 of which 62.6\% is spent on (de)compressing columns.

\paratitle{Other Major Challenges} 
Like with single-instance
design, we find that maintaining data consistency for multiple instances without compromising
performance isolation is very challenging. Updates from transactions are frequently shipped
and applied to analytics replicas while analytical queries run. 
As a result, multiple-instance-based systems suffer from the same consistency drawbacks
that we observe for single-instance-based systems in Section~\ref{sec:bkgd:single}.
 Another major challenge we find is the limited performance isolation. While separate
 instances provide partial performance isolation, as transactions and analytics do not compete for
 the same copy of data, they still share underlying hardware resources such as CPU cores and 
 the memory system. As we discuss in Section~\ref{sec:bkgd:single},
 analytics workloads, as well as data freshness and consistency mechanisms, generate 
 a large amount of data movement and take many cycles.
As a result, multiple-instance designs also suffer from limited performance isolation.

\vspace{3pt}
We conclude that neither single- nor multiple-instance HTAP systems meet all of
our desired HTAP properties. 
We therefore need a system that can avoid \sgdel{shared }resource contention and alleviate
the \sgdel{high }data movement costs incurred for HTAP.

%% file: sections/proposal.tex

\section{\texorpdfstring{\MakeUppercase\titleShort}{\titleShort}}
\label{sec:proposal}

\sgmod{Our goal in this work is to design an HTAP system that can meet 
all of the desired HTAP properties, by avoiding the challenges that we
identify for state-of-the-art HTAP systems in Section~\ref{sec:bkgd:motiv}.
To this end, we propose \titleLong (\titleShort).
\titleShort}{We propose \titleShort, which} divides the HTAP system into multiple \emph{islands}.
Each island includes (1)~a replica of data whose layout is optimized for a specific workload, 
(2)~an optimized execution engine, and (3)~a set of hardware \sgmod{resources (e.g., computation units, memory).}{resources.}
 \titleShort has two types of islands: 
(1)~a \emph{transactional island}, and
(2)~an \emph{analytical island}. 
\rev{To avoid the data movement and interference
challenges that other multiple-instance-based HTAP systems face (see
Section~\ref{sec:bkgd:motiv}),}
we propose to equip each analytical island with
(1)~\emph{in-memory hardware}; and
(2)~co-designed algorithms and hardware for the analytical execution engine,
\emph{update propagation}, and 
\emph{consistency}.

\begin{figure}[t]
    \vspace{-5pt}
    \centering
        \centering
        \includegraphics[width=\linewidth]{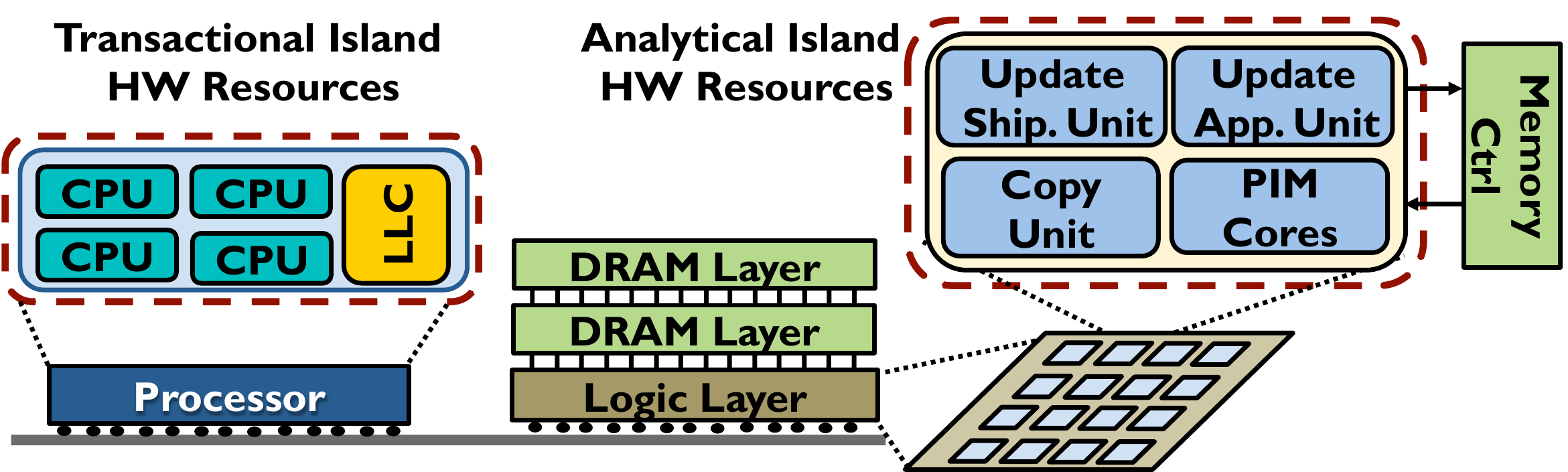}%
\vspace{-22pt}
    \caption{High-level organization of \titleShort hardware.}
    \label{fig:high-level-hw}
    \vspace{-15pt}
\end{figure}

\titleShort is a framework that can be applied to many different combinations
of transactional and analytical workloads.
 In this work, we focus on designing an instance of \titleShort 
 that supports relational transactional and analytical workloads.\footnote{Note that our proposed techniques
 can be applied to other types of analytical workloads (e.g., graphs, machine learning) as well.}
Figure~\ref{fig:high-level-hw} shows the hardware for our chosen implementation,
which includes 
one transactional island and one analytical island, and is equipped with
a 3D-stacked memory similar to the Hybrid Memory Cube (HMC)~\cite{hmcspec2},
where multiple vertically-stacked DRAM cell layers are connected with a 
\emph{logic layer} using thousands of \emph{through-silicon vias} (TSVs). 
An HMC chip is split up into multiple \emph{vaults}, where each vault corresponds to 
a vertical slice of the memory and logic layer.
The transactional island uses an execution engine similar
 to conventional transactional engines~\cite{dbx1000,kallman2008h} to execute a relational 
 transactional workload. The transactional island is equipped with conventional multicore CPUs
 and multi-level caches, as transactional queries have
 short execution times, are latency-sensitive, and have cache-friendly access patterns~\cite{conda}. 
Inside each vault's portion of the logic layer in memory, 
we add hardware for the analytical island, including
the update propagation mechanism (consisting of the 
\emph{update shipping} and \emph{update application} units), the consistency
 mechanism (\emph{copy units}), and the 
analytical execution engine (simple programmable in-order PIM cores).%
\sgdel{In the next three sections, we discuss the detailed design of the
update propagation mechanism (Section~\ref{sec:proposal:update-propagation}),
consistency mechanism (Section~\ref{sec:proposal:consistency}), and 
analytical execution engine (Section~\ref{sec:proposal:analytic-engine}).}

\sgiv{To address potential capacity issues and accommodate larger data, 
\titleShort can extend across multiple memory stacks.  We evaluate
\titleShort with multiple stacks in Section~\ref{sec:eval:multiple}.}

%% file: sections/update-propagation.tex

\section{Update Propagation Mechanism}
\label{sec:proposal:update-propagation}

We design \sgii{a new two-part} update propagation mechanism to overcome the high costs
\sgii{of analytical replica updates in state-of-the-art HTAP systems}.
The \emph{update shipping unit} gathers updates from the transactional island,
finds the target location in the analytical island, and frequently 
pushes these updates to the analytical island.
The \emph{update application unit} receives these updates,
converts the updates from the transactional replica data format to the
analytical replica data format, and applies the update to the
analytical replica.

\subsection{Update Shipping}
\label{sec:proposal:update-propagation:update-shipping}

\paratitle{Algorithm} 
\rev{Our update shipping mechanism includes three major stages. For each thread in} the transactional engine, \titleShort stores an ordered \emph{update log}
for the queries performed by the thread.
Each update log entry contains four fields: 
(1)~a commit ID (a timestamp used to track the total order of all
updates across threads), (2)~the type
 of the update (insert, delete, modify), (3)~the updated data, and (4)~a record key (e.g., pair of
 row-ID and column-ID) that links this particular update to a column in
the analytic replica. 
 The update shipping process is triggered when total number of pending updates reaches the final log
 capacity, which we set to 1024 entries (see Section~\ref{sec:proposal:update-propagation:update-application}).
The first stage is to scan the per-thread update logs,
and merge them into a single \emph{final log}, where all updates are sorted by the commit ID.

 The second \rev{stage} is to find the location of the corresponding column (in the analytical replica) 
 associated with each update log entry.
 \rev{We observe that this stage is one of the major bottlenecks of update shipping, because the
 fields in each tuple in the transactional island are distributed across different columns
 in the analytical island. Since the column size is typically very large, 
 finding the location of each update is a very time-consuming process. To overcome this,}
we maintain a hash index of data on the (column,row) key, and
 use that to find the corresponding column for each update
 in the final log. 
\rev{We use the modulo operation as the hash function. We size our hash table based on
 the column partition size. Similar to conventional analytical DBMSs, we can
 use soft partitioning~\cite{batchdb, morsel, scaling-up-c-store-numa} to address scalability
 issues when the column size increases beyond a certain point. Thus, the hash table
 size does not scale with column size.} 
This stage contains a buffer for each column in the analytical island,
and we add each update from the final log to its corresponding column buffer.

The final stage is to ship these buffers to each
 column in the analytical replica. 

\paratitle{Hardware}
\rev{We find that despite our best efforts to minimize overheads, our algorithm
has three major bottlenecks that
 keep it from meeting data freshness and performance isolation requirements:}
(1)~the scan and merge \rev{operation} in stage~1,
(2)~hash index lookups in stage~2, and
(3)~transferring the column buffer contents to the analytical islands in stage~3.
\rev{These primitives generate a large amount of data movement and account for 87.2\% of
 our algorithm's execution time.}
To avoid these bottlenecks, 
we \sgii{design a new hardware accelerator, 
called the \emph{update shipping unit}, \rev{that speeds up the key primitives of the update shipping algorithm.}
We add \rev{this accelerator} to each of
\titleShort's in-memory analytical islands}.

\begin{figure}[h]
    \vspace{-4pt}
    \centering
    \includegraphics[width=\linewidth]{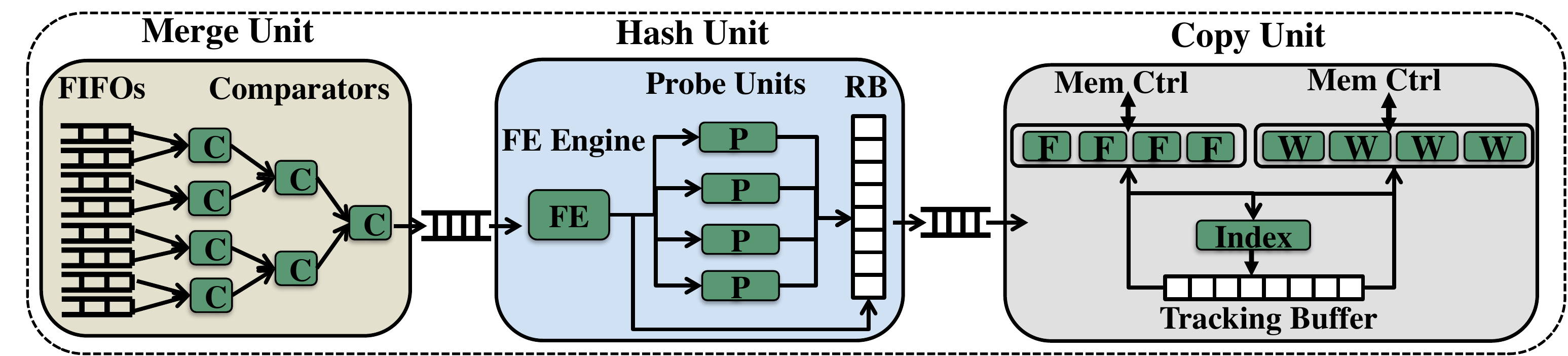}%
    \vspace{-20pt}
    \caption{\rev{Update shipping unit architecture.}}
    \label{fig:update-shipping-hw}
    \vspace{-5pt}
\end{figure}

\rev{Figure~\ref{fig:update-shipping-hw} shows the high-level architecture of our in-memory 
update shipping unit.}
The update shipping unit consists of three building 
blocks: (1)~a merge unit, (2)~a hash lookup unit, and (3)~a copy unit. The merge unit 
\sgii{consists of 8 FIFO input queues, where each input queue corresponds to a sorted update log.}
\rev{Each input queue can hold up to 128 updates, which are streamed from DRAM. 
The merge unit finds the oldest entry among the input queue heads,
using a 3-level comparator tree, and adds it to
the tail of the final log (a ninth FIFO queue). The final log is then sent to the hash
 unit to determine the target column for each update.}

\rev{For our hash unit, we start with a preliminary design that
includes a \emph{probe
 unit}, a simple finite state machine controller that takes the address (for the key), computes the hash function to find the
 corresponding bucket in memory, and traverses the linked list of keys in the bucket.
We find that having a single probe unit 
does not achieve our expected performance because (1)~we cannot fully
 exploit the bandwidth of 3D-stacked memory;
 and (2)~each lookup typically
 includes several pointer chasing operations
that often leave the probe unit idle. As a result, we need to perform multiple hash lookups in parallel
 at each step. However, the challenge 
     is that updates need to be handed to the copy unit in the same commit order they are inserted
 into the final update log

To address these challenges, we design the hash unit shown in Figure~\ref{fig:update-shipping-hw}. 
Its key idea is to (1)~decouple hash computation
 and bucket address generation from the actual bucket access/traversal, to allow for concurrent
 operations; and (2)~use a small reorder buffer to track
 in-flight hash lookups, and maintain commit order for completed lookups that are sent to the copy engine.
 We introduce a front-end engine that fetches the keys from the final update
 log, computes the hash function,
and sends the key address to probe units. 
 The front-end engine allocates an entry (with the bucket address and a ready bit) for each lookup in the reorder buffer.
We employ multiple probe
 units (we find that 4 strikes a balance between parallelism and area overhead), with each taking a bucket address
 and accessing DRAM to traverse the linked list.}

We describe our copy engine in Section~\ref{sec:proposal:consistency}.
Our analysis shows that the total area of our update shipping unit is \SI{0.25}{\milli\meter\squared}.

\subsection{Update Application}
\label{sec:proposal:update-propagation:update-application}

Similar to other relational analytical DBMSs, our
analytical engine uses the DSM data layout and 
\emph{dictionary encoding}~\cite{c-store,update-c-store,sap-hana,compression-gpu,numa-scan,dict-compression,compression-c-store}. 
With dictionary encoding, each column in a table is transformed into a compressed column using
encoded fixed-length integer values, and a dictionary stores a sorted mapping of
real values to encoded values. 
As we discuss in Section~\ref{sec:bkgd:multiple}, 
the layout conversion (our transactional island uses NSM)
and column compression 
make update application process challenging. 
\sgii{We design a new update application mechanism for \titleShort 
that uses hardware--software co-design to address these challenges.}

\paratitle{Algorithm} 
\sgii{We first discuss an initial algorithm that we develop for update application.}
We assume each column has \emph{n} entries, and that we have \emph{m} \sgdel{number of }updates.
First, the algorithm decompresses the encoded column by scanning the column and
looking up in the dictionary to decode each item. 
This requires \emph{n random accesses} to the dictionary. 
Second, the algorithm 
applies updates to the
decoded column one by one.
Third, 
it constructs a new dictionary, by sorting the updated column 
and calculating the number of fixed-length integer bits required to encode the
sorted column.
Dictionary construction is computationally expensive ($\mathcal{O}((n+m)\log{}(n+m))$)
because we need to sort the entire column.
Finally, the algorithm compresses the new column
using our \sgmod{newly-constructed}{new} dictionary.
While entry decoding happens in constant time, encoding requires a
logarithmic complexity search through the dictionary (since the dictionary is sorted).

This \sgii{initial} algorithm is memory intensive (Steps~1, 2, 4)
and computationally expensive (Step~3). Having hardware support is \emph{critical} to enable
low-latency update application and performance isolation. 
While \sgmod{processing-in-memory (PIM)}{PIM} may be able to help,
our \sgii{initial} algorithm is not well-suited for PIM for two reasons,
and we \sgii{optimize} the algorithm to address both.

\emph{Optimization~1: Two-Stage Dictionary Construction.}
\sgii{We eliminate column sorting from Step~3, as it is
computationally expensive.}
Prior work~\cite{pim-sort, q100} shows that to efficiently sort more than 1024 values in hardware,
we should provide a hardware partitioner to split the values into
multiple chunks, and then use a sorter unit to sort chunks one at a time.
This requires an area of \SI{1.13}{\milli\meter\squared}~\cite{pim-sort, q100}.
\sgii{Unfortunately, since tables can have millions of entries~\cite{update-c-store},
we would need multiple sorter units to construct a new dictionary,
easily exceeding the total area budget of \SI{4.4}{\milli\meter\squared}
per vault~\cite{google-pim,tetris,mondrian}.}

\sgii{To eliminate column sorting, we sort only the dictionary, 
leveraging} the fact that
(1)~the existing dictionary is already sorted, and
(2)~the new updates are limited to 1024~values.
\sgmod{As a result, our}{Our} optimized algorithm initially builds a sorted dictionary for
only the updates, which requires a single hardware \sgmod{sorter.
A 1024-value bitonic sorter requires a much smaller area of only
\SI{0.18}{\milli\meter\squared}~\cite{q100}.}{sorter (a 1024-value bitonic sorter 
with an area of only \SI{0.18}{\milli\meter\squared}~\cite{q100}).}
Once the update dictionary is constructed, we now have two sorted dictionaries:
the old dictionary and the update dictionary.
We merge these into a single dictionary using a linear scan ($\mathcal{O}(n+m)$),
and then calculate the number of bits required to encode the new dictionary.

\emph{Optimization~2: Reducing Random Accesses.}
To reduce \sgii{the algorithm's memory intensity (which is a result of random lookups),}
we maintain a hash index that links the old encoded value in a column to the new encoded
value. 
This avoids the need to decompress the column and add updates, eliminating
data movement and random accesses for Steps~1 and 2, while reducing the
number of dictionary lookups required 
\sgii{for Step~4.}
The only remaining random accesses are for Step~4, which decrease from
$\mathcal{O}((n+m)\log{}(n+m))$
to 
$\mathcal{O}(n+m)$.

We now describe our optimized algorithm.
We first sort the updates to construct the update dictionary.
We then merge the old dictionary and the update dictionary to construct the new dictionary
and hash index.
Finally,
we use the index and the new dictionary to find the new encoded
value for each entry in the column.

\paratitle{Hardware} 
We \sgii{design a hardware implementation of our optimized algorithm,
called the \emph{update application unit}, and add it to each in-memory
analytical island.}
The unit consists of three building blocks: a \emph{sort unit}, a \emph{hash lookup unit}, and
a \emph{scan/merge unit}. 
\sgii{Our sort unit uses} a 1024-value bitonic sorter,
\rev{whose basic building block is a network of comparators. These comparators are
 used to form \emph{bitonic sequences}, sequences where the first half of the sequence is monotonically
 increasing and the second half is decreasing.
The hash lookup uses a simpler version of the component that we designed for update shipping.
The simplified version does not use a reorder buffer, as there is no dependency between hash lookups
 for update application. We use the same number of hash units
(empirically set to 4), each corresponding to one index structure, to parallelize the compression
 process.} For the merge unit, we use a similar design from our update shipping unit. 
Our analysis shows that the total area of our update
application unit is \SI{0.4}{\milli\meter\squared}.

%% file: sections/consistency.tex

\section{Consistency Mechanism}
\label{sec:proposal:consistency}

\sgii{We design a new consistency mechanism for \titleShort in order} to deliver all of the 
desired properties of an HTAP system (Section~\ref{sec:bkgd:requirements}).
Our consistency mechanism must not compromise either the throughput of
analytical queries or the rate at which updates are applied.  This sets two
\am{requirements} for our mechanism:
(1)~analytics must be able to run all of the time without slowdowns, 
to satisfy the performance isolation property; and
(2)~the update application process should not be blocked by long-running
analytical queries, to satisfy the data freshness property.
This means that our mechanism needs a way to allow analytical queries
to run concurrently with updates, without incurring the long chain read
overheads of similar mechanisms such as MVCC (see Section~\ref{sec:bkgd:single}).

\paratitle{Algorithm} 
Our mechanism relies on a combination of snapshotting~\cite{hyper} and versioning~\cite{mvcc} 
to provide snapshot isolation~\cite{serializable-isolation,isolation-level} for analytics. 
Our consistency
mechanism is based on two key observations: (1)~updates are applied at a column
granularity, and (2)~snapshotting a column is cost-effective using PIM logic. We assume
that for each column, there is a chain of snapshots where each chain entry corresponds to
a version of this column. Unlike chains in MVCC, each version is associated with a
column, not a tuple. 

We adopt a lazy approach (late materialization~\cite{abadi2007materialization}), where \titleShort does not create a snapshot
every time a column is updated. 
Instead, \sgii{on a column update}, \rev{\titleShort marks} the column as dirty, indicating that
the snapshot chain does not contain the most recent version of the column data.
\sgii{When an analytical query arrives \titleShort checks the column \rev{metadata}, and 
creates a new snapshot only if \am{(1)~any of the columns are dirty (similar to Hyper~\cite{hyper}), and (2)~no 
\sgiii{current} snapshot exists for the same column (\sgiii{we let multiple queries share a single snapshot}).}
During snapshotting, \titleShort}
updates the head of the snapshot chain with the new value, and \rev{marks} the column as clean.
This \sgii{provides two benefits.
First, the analytical query avoids the chain traversals and timestamp comparisons performed in MVCC,
as the query only needs to access the head of the chain at the time of the snapshot.
Second, \rev{\titleShort uses \am{a simple yet} efficient garbage collection:} 
 when an analytical query finishes,
snapshots no longer in use by any query are deleted 
(aside from the head of the chain).}

 To guarantee high data freshness (\emph{second requirement}), our
 consistency mechanism always lets transactional updates directly
update the main replica 
\rev{using our two-phase update application algorithm
(Section~\ref{sec:proposal:update-propagation:update-application}).
In Phase~1, the algorithm constructs a new
dictionary and a new column.
In Phase~2, the algorithm atomically updates the main replica with pointers to the
new column and dictionary.}

\paratitle{Hardware}
\am{\sgiii{Our algorithm's success at} satisfying
 the \sgiii{first requirement for a consistency mechanism (i.e., no slowdown for
analytics) relies heavily on its ability to perform fast memory copies
to minimize the snapshotting latency.}
\sgiii{Therefore, we} \sgii{add a custom copy unit to
each of \titleShort's in-memory analytical \sgiii{islands.}}
\sgiii{We have two design
 goals for the unit.} First, the accelerator needs to be able to issue multiple memory accesses concurrently.
This is because (1)~we are designing the copy engine for an
arbitrarily-sized memory region (e.g., a column), which is often larger than 
the memory access granularity per vault (8--16B) in an HMC-like \sgiii{memory}; and (2)~we 
want to fully exploit the internal bandwidth of 3D-stacked memory. 
Second, \sgiii{when a read for a copy completes, the accelerator
should immediately initiate the write.}}

\am{We design \sgiii{our} copy unit (Figure~\ref{fig:update-shipping-hw}) to satisfy \sgiii{both} design goals. To issue multiple memory accesses concurrently,
we leverage the observation that these memory accesses are independent. We
use multiple \sgiii{fetch and writeback units, which can read from or write to
source/destination regions in parallel.}
To satisfy the second design goal, we need to track outstanding reads, as
 they may come back from memory \sgiii{out of order}. 
Similar to prior work on accelerating \texttt{memcpy}~\cite{memcopy-accelerator},
\sgiii{we use a \emph{tracking buffer} in our copy unit.  The buffer allocates} an entry for each read issued to memory,
where an entry contains a memory address and a ready bit.
Once a read completes, we find its corresponding entry in \sgiii{the buffer} and
set its ready bit \sgiii{to trigger the write}.}

\am{\sgiii{We find that the buffer lookup limits the performance of the copy unit, as 
each lookup results in a full buffer scan,} and multiple fetch units \sgiii{perform
lookups concurrently (generating high contention)}. To \sgiii{alleviate this}, we design a hash index based on
the memory address to determine the location of a read in the buffer. \sgiii{We make use a 
similar design as the hash lookup unit in our update shipping unit.}}

%% file: sections/analytic-engine.tex

\section{Analytical Engine}
\label{sec:proposal:analytic-engine}

The analytical execution engine performs the analytical queries.
When a query arrives, the engine parses the query and generates \rev{an algebraic} query plan
consisting of physical operators (e.g., scan, filter, join).
\rev{In the query plan, operators are arranged in a tree where data flows from the bottom nodes
(leaves) toward the root, and the result of the query is stored in the root. The analytical execution 
engine employs the top-down Volcano (Iterator) execution model~\cite{volcano, volcano2} to traverse
 the tree and execute operators while respecting dependencies between operators. Analytical queries
 typically exhibit a high degree of both intra- and inter-query parallelism~\cite{qtm,morsel,scaling-up-c-store-numa}. To exploit this,
the engine decomposes a query into multiple tasks, each being a sequence of one or more operators.}
The engine \am{(task scheduler)} then schedules the tasks, often in a way that
executes multiple independent tasks in parallel.

Efficient analytical query execution strongly depends on (1)~data layout and data placement,
(2)~the task scheduling policy, and (3)~how each physical operator is executed.
Like prior works~\cite{mondrian, JAFAR}, we find that the \rev{execution of} 
physical operators of analytical queries 
significantly benefit from PIM. 
However, without a HTAP-aware and PIM-aware data placement strategy and task scheduler,
PIM logic for operators alone cannot provide significant throughput improvements.

We design \sgii{a new} analytical execution engine based on the characteristics of our in-memory hardware.
As we discuss in Section~\ref{sec:proposal}, \titleShort uses a 3D-stacked
memory that contains multiple vaults.
Each vault
(1)~provides only a fraction (e.g., 8 GB/s) of the total bandwidth available
in a 3D-stacked memory,
(2)~has limited power and area budgets for PIM logic, and
(3)~can access its own data faster than it can access data stored in other vaults
(which take place through a vault-to-vault interconnect).
We take these limitations into account as we design our data placement mechanism
and task scheduler.

\subsection{Data Placement}
\label{sec:proposal:analytic-engine:placement}

We evaluate three different data placement strategies for \titleShort.
Our analytical engine uses the DSM layout to store data, and
makes use of dictionary encoding~\cite{dict-compression} for column compression. 
Our three strategies affect which vaults the compressed DSM columns and dictionary are
stored in.

\paratitle{Strategy~1: Store the Entire Column (with Dictionary) in One Vault}
This strategy has two major benefits. First, both the dictionary lookup and column
access are to the local vault, which 
improves analytical query throughput. Second, it simplifies the update application
process (Section~\ref{sec:proposal:update-propagation:update-application}), as
the lack of remote accesses avoids the need to synchronize updates between 
multiple update applications units (as each vault has its own unit).

However, this data placement strategy forces us to service tasks
that access a particular column using (1)~highly-constrained
PIM logic (given the budget of a single vault), 
and (2)~only the bandwidth available to one vault. 
This significantly degrades throughput, 
as it becomes challenging to benefit from
intra-query parallelism with only one highly-constrained
set of logic available for a query.

\paratitle{Strategy~2: Partition Columns Across All Vaults in a Chip}
To address the challenges of Strategy~1, we can partition each column and distribute it 
across all of the vaults in the 3D-stacked memory chip (i.e., a \emph{cube}). 
This approach allows us to (1)~exploit the
entire internal bandwidth of the 3D-stacked memory, and (2)~use all
of the available PIM logic to service
each query. Note that unlike partitioning the column, partitioning the
dictionary across all vaults is challenging because the dictionary is sorted,
forcing us to scan the entire column and find the corresponding dictionary entries
 for each column entry. 

However, Strategy~2 suffers from two major drawbacks. First, it makes the update application 
(Section~\ref{sec:proposal:update-propagation:update-application}) significantly
challenging. To perform update application under Strategy~2, we need to (1)~perform many 
remote accesses to gather all of the column partitions, (2)~update the column, and then 
(3)~%
perform many remote accesses to scatter the updated column partitions back across the vaults. 
Given the high frequency of update application in HTAP workloads, 
this gather/scatter
significantly increases the latency of update application and intra-cube traffic.
Second, we need to perform several remote accesses to collect sub-results from each partition
and aggregate them, which reduces the throughput.

\paratitle{Strategy~3: Partition Columns Across a Group of Vaults} 
To overcome the challenges of Strategies~1 and 2, we propose a hybrid strategy
where we create small \emph{vault groups}, consisting of a fixed number of vaults,
and partition a column across the vaults in a vault group.
For a group with \emph{v} vaults, this allows us to increase the aggregate 
bandwidth 
for servicing each query by \emph{v} times,
and provides up to \emph{v} times the power and area for PIM logic.
\sgii{The number of vaults per group is critical for efficiency:} too many vaults can
complicate the update application process, while not enough vaults can degrade throughput.
We empirically find that four vaults per group strikes a good balance.

While the hybrid strategy reduces the cost of update application compared to Strategy~2,
it still needs to perform remote accesses within each vault group. 
To \sgii{overcome this}, we leverage an observation from prior work~\cite{update-c-store} that
the majority of columns have only a limited (up to 32)
number of distinct values. This means that 
the entire dictionary 
incurs negligible storage overhead (\textasciitilde 2 KB). 
\sgii{To avoid remote dictionary accesses during update application, Strategy~3
keeps a copy of the dictionary in each vault.}
Such an approach is \sgii{significantly costlier} under Strategy~2, as for a given column size,
the number of dictionary copies scales linearly with the number of column partitions,
\sgii{which is particularly problematic in a system with multiple memory stacks.}

\sgii{\titleShort makes use of Strategy~3 for data placement.}

\subsection{Scheduler}
\label{sec:proposal:analytic-engine:scheduler}

\sgii{\titleShort's} task scheduler 
plays a key role in (1)~exploiting inter- and intra-query 
parallelism, and (2)~efficiently utilizing hardware resources.
For each query, the scheduler (1)~decides how many tasks to create,  
(2)~finds how to map these tasks to the available \sgdel{hardware }resources (\emph{PIM threads}), and
(3)~guarantees that dependent tasks are executed in order. 
\rev{We first design a basic scheduler heuristic that generates tasks (statically at compile time)
by disassembling the operators of the query plan into operator instances (i.e., an
 invocation of a physical operator on some subset of the input tuples) based on (1)~which vault groups the input tuples reside in; and (2)~the number of
 available PIM threads in each vault group,
 which determines the number of tasks generated. 
The scheduler inserts tasks into a global work queue
in an order that preserves dependencies between operators,
monitors the progress of PIM threads, and
assigns each task to a free thread (push-based assignment). 

However, we find that this heuristic is not optimized for PIM, and leads to sub-optimal performance 
due to three reasons. First, the heuristic requires a dedicated
 runtime component to monitor and assign tasks.
The runtime component must be executed on a general-purpose PIM core,
either requiring another core (difficult given limited area/power budgets) or
preempting a PIM thread on an existing core (which hurts performance).
Second, the heuristic's static mapping is limited to using only the resources
available within a single vault group, which can lead to performance issues
for queries that operate on very large columns.
Third, this heuristic is vulnerable to load imbalance, as
some PIM threads might finish their tasks sooner and wait idly for straggling threads. 

We optimize our heuristic to address these challenges. First, 
we design a pull-based task assignment strategy, where PIM threads cooperatively
 pull tasks from the task queue at runtime. This eliminates the need for a
 runtime component (first challenge) and allows PIM thread to 
 dynamically load balance (third challenge). 
To this end, we introduce a local task queue for each vault group. Each PIM thread looks into its
  own local task queue to retrieve its next task. Second, we optimize the heuristic to 
  allow for finer-grained tasks. 
 Instead of mapping tasks statically,
we partition input tuples into fixed-size segments (i.e., 1000 tuples) and
  create an operator instance for each partition. The
  scheduler then generates tasks for these operator instances
 and inserts them into corresponding task queues (where those tuple segments reside).
 The greater number of tasks increases opportunities for load balancing. 
Finally, we optimize the heuristic to allow a PIM thread to steal tasks
 from a remote vault if its local queue is empty. This enables us to potentially use all available
 PIM threads to execute tasks, regardless of the data distribution (second challenge). Each PIM
 thread first attempts to steal tasks from other PIM threads in its own vault group, because the
 thread already has a local copy of the full dictionary in its vault, and needs remote vault
 accesses only for the column partition. If there is no task to steal in its vault group, the PIM thread
 attempts to steal a task from a remote vault group.}

\subsection{Hardware Design}
\label{sec:proposal:analytic-engine:hardware}

Given area and power constraints, it can be difficult to add enough PIM logic to each vault to 
saturate the available vault bandwidth~\cite{mondrian}. Mondrian~\cite{mondrian}
attempts to change the access pattern from random to sequential, allowing PIM threads to use stream buffers
to increase bandwidth utilization. 
With our new data placement strategy and scheduler, we instead expose greater intra-query
parallelism, \rev{and use simple programmable in-order PIM cores to exploit the available vault bandwidth.
We add} four PIM cores
to each vault, where the cores are similar to those in prior work~\cite{Mingyu:PACT,mondrian}.
\change{We run} a PIM thread on each core, and \change{we use} these cores to execute
the scheduler and other parts of the analytical engine (e.g., query parser).

\rev{We find that our optimized heuristic significantly increases data sharing
    between PIM threads. This is because within each vault group, all 16 PIM threads
        access the same local task queue, and must synchronize their accesses.
        The problem worsens when other PIM threads attempt to steal tasks from remote vault groups, especially for highly-skewed workloads.
        To avoid excessive accesses to DRAM and let PIM
        threads share data efficiently, we implement a simple fine-grained coherence technique, which uses
        a local PIM-side directory in the logic layer to implement a low-overhead coherence protocol.}

%% file: sections/methodology.tex

\section{Methodology}
\label{sec:methodology}

We use and heavily extend state-of-the-art transactional and analytical engines to 
implement various single- and multiple-instance HTAP configurations.
We use DBx1000~\cite{dbx1000,dbx1000-github} as the starting point for
our transactional engine, and we implement an in-house analytical engine
similar to C-store~\cite{c-store}.
Our analytical engine supports key physical
operators for relational analytics (select, filter, aggregate and join), and
supports both NSM and DSM layouts, and dictionary encoding~\cite{dict-compression,compression-c-store,numa-scan}.
For consistency, we implement both snapshotting (similar to software 
snapshotting~\cite{software-snapshotting}, \sgii{and with snapshots taken 
only when dirty data exists}) and MVCC (adopted from DBx1000~\cite{dbx1000}). 

Our baseline single-instance HTAP system
stores the single data replica in main memory.
Each transactional query randomly performs reads or writes on a few randomly-chosen tuples from a randomly-chosen
table. Each analytical query uses select and join on randomly-chosen tables and columns. 
Our baseline multiple-instance HTAP system models a similar system as our single-instance baseline,
but provides the transactional and analytical engines with separate replicas
(using the NSM layout for transactions, and DSM with dictionary encoding for analytics).
Across all baselines, we have 4 transactional
and 4 analytical worker threads. 

We simulate \titleShort using gem5~\cite{GEM5},
integrated with DRAMSim2~\cite{DRAMSim2} to model 
an HMC-like 3D-stacked DRAM~\cite{hmcspec2}.
Table~\ref{tbl:config} shows our system configuration.
For the analytical island, 
each vault of our 3D-stacked memory contains
four PIM cores and three fixed-function accelerators (update shipping unit,
update application unit, copy unit).
For the PIM core, we model a core
similar to the ARM Cortex-A7~\cite{cortex-a7}.

\begin{table}[h]
    \small
    \vspace{-8pt}
    \renewcommand{\arraystretch}{0.5}
    \setlength{\aboverulesep}{1pt}
    \setlength{\belowrulesep}{2pt}
    \centering
    \setlength{\tabcolsep}{.5em}
    \begin{tabular}{ll}
        \toprule
        \emph{Processor} &  
        4 OoO cores, \change{each with 2 HW threads}, 8-wide issue; \\ 
        \emph{(Transactional} &     
        \emph{L1 I/D Caches}: \SI{64}{\kilo\byte} private, 4-way assoc.; \emph{L2 Cache:} \\
	\emph{Island)} & \SI{8}{\mega\byte} shared, 8-way assoc.; \emph{Coherence}: MESI \\
        \midrule
        \emph{PIM Core} & 
           4 in-order cores per vault, 2-wide issue, \\ 
        & \emph{L1 I/D Caches}: \SI{32}{\kilo\byte} private, 4-way assoc. \\
        \midrule
        \emph{3D-Stacked} & 
              \SI{4}{\giga\byte} cube, 16 vaults per cube; \emph{Internal Bandwidth:} \\
              \emph{Memory} &  \SI{256}{\giga\byte}/s; \emph{Off-Chip Channel Bandwidth:} \SI{32}{\giga\byte}/s \\
        \bottomrule
    \end{tabular}%
 \vspace{-10pt}
    \caption{Evaluated system configuration.}
    \label{tbl:config}%
   \vspace{-8pt}
\end{table}

\change{\textbf{Area and Energy.}
Our four PIM cores require
\SI{1.8}{\milli\meter\squared}, based on the
Cortex-A7 (\SI{0.45}{\milli\meter\squared} each)~\cite{cortex-a7}.
We use Calypto Catapult to determine the area of the accelerators for
 a 22nm process: \SI{0.7}{\milli\meter\squared} for the update propagation units
 and \SI{0.2}{\milli\meter\squared} for the in-memory copy unit for our consistency mechanism. This
 brings \titleShort's total to \SI{2.7}{\milli\meter\squared} per vault.}
We model system energy
similar to prior work~\cite{jeddeloh2012hybrid, google-pim, conda}, which sums the energy consumed
by the CPU cores, all caches (modeled using CACTI-P 6.5~\cite{CACTI}),
DRAM, and all on-chip and off-chip interconnects.

%% file: sections/evaluation.tex

\section{Evaluation}
\label{sec:eval}

\sgdel{We first evaluate the three major components of our proposal:
(1)~update propagation, (2)~our consistency mechanism, and (3)~our analytical
engine. We then perform an end-to-end system evaluation.}

\subsection{End-to-End System Analysis}

\begin{figure}[h]
    \vspace{-10pt}
    \centering
        \centering
        \includegraphics[width=\linewidth]{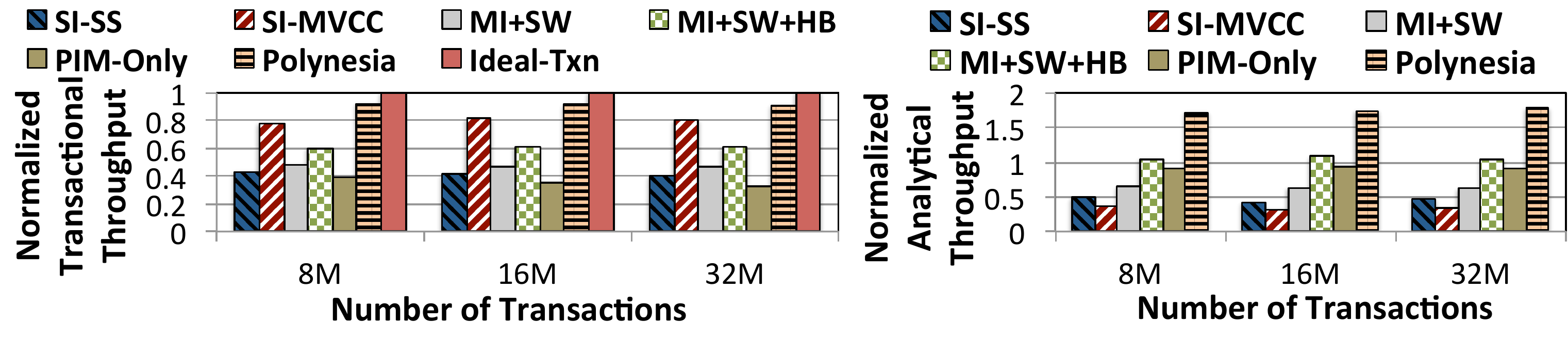}%
\vspace{-25pt}
    \caption{Normalized transactional (left) \change{and analytical (right)} throughput for end-to-end HTAP systems.}
    \label{fig:eval-end-to-end}
    \vspace{-10pt}
\end{figure}

Figure~\ref{fig:eval-end-to-end} \change{(left)} shows the transactional throughput 
\change{for \rev{six} DBMSs:
(1)~Single-Instance-Snapshot (\rev{\emph{SI-SS}});
(2)~Single-Instance-MVCC (\emph{SI-MVCC});
(3)~\emph{MI+SW}, an improved version of
 Multiple-Instance that includes all of our software optimizations for \titleShort (except those
 specifically targeted for PIM);
(4)~\emph{MI+SW+HB}, a hypothetical version of 
MI+SW with 8x bandwidth (256 GB/s), equal to the
 internal bandwidth of HBM; \rev{(5)~\emph{PIM-Only}, a hypothetical version of 
MI+SW which uses general-purpose PIM cores to run both transactional and analytical workloads; and
(6)}~\emph{\titleShort}, our full hardware--software proposal.}
We normalize throughput to an ideal transaction-only DBMS
(\emph{Ideal-Txn}) \change{for each transaction count}.
\change{Of the single-instance DBMSs, SI-MVCC performs best, coming within 20.0\% of the throughput of Ideal-Txn on average.
Its use of MVCC over snapshotting overcomes the high performance penalties incurred by
SI-SS.
For the two software-only multiple instance DBMSs (MI+SW and MI+SW+HB), 
despite our enhancements, both fall significantly short of SI-MVCC due to their
lack of performance isolation and, in the case of MI+SW,
the overhead of update propagation.}
\change{MI+SW+HB, even with its higher available bandwidth, cannot data movement or contention 
on shared resources. As a result, its transactional throughput is still 41.2\% lower than Ideal-Txn.
\rev{PIM-only significantly hurts transactional throughput (by 67.6\% vs. Ideal-Txn),
 and even performs 7.6\% worse than SI-SS.}
\titleShort improves the average throughput by 51.0\% over MI+SW+HB, 
and by 14.6\% over SI-MVCC, because it
(1)~uses custom PIM logic for analytics along with its update propagation
 and consistency mechanisms to significantly reduce contention, and (2)~reduces off-chip
bandwidth contention by reducing data movement.}
\sgii{As a result, \titleShort comes within 8.4\% of Ideal-Txn.}

\change{Figure~\ref{fig:eval-end-to-end} (right) shows the analytical throughput across the same DBMSs. 
We normalize throughput at each transaction count to a baseline where analytics are running alone on
 our multicore system. We see that while SI-MVCC is the best software-only DBMS for transactional throughput,
it degrades analytical throughput by 63.2\% compared to the analytics baseline, due to its lack of
 workload-specific optimizations and poor consistency mechanism (MVCC). Neither of these problems can be
 addressed by providing higher bandwidth. MI+SW+HB is the best
 software-only HTAP DBMS for analytics, because it provides workload-specific optimizations, 
but it still loses 35.3\% of the analytical throughput of the baseline. 
MI+SW+HB improves throughput by 41.2\% over MI+SW but still suffers from
 resource contention due to update propagation and the consistency mechanism.
\rev{PIM-Only performs similar to MI+SW+HB but reduces throughput by 11.4\% compared to that, as it suffers 
from resource contention caused by co-running transactional queries.}
\titleShort \emph{improves} over the baseline by 63.8\%,
by eliminating data movement, having low-latency accesses, and using
custom logic for update propagation and consistency.}

\sgii{Overall, averaged across all transaction counts in Figure~\ref{fig:eval-end-to-end}, 
\titleShort has a higher transactional throughput
(2.20X over SI-SS, 1.15X over SI-MVCC, and 1.94X over MI+SW;
mean of 1.70X), \emph{and} a higher analytical throughput
(3.78X over SI-SS, 5.04X over SI-MVCC, and 2.76X over MI+SW;
mean of 3.74X).}

\rev{\paratitle{Real Workload Analysis}. To model more complex queries, we evaluate \titleShort using a mixed workload 
from TPC-C~\cite{TPC-C} (for our transactional workload) and TPC-H~\cite{TPC-H} (for our analytical workload). TPC-C's 
schema includes nine relations (tables) that simulate an order processing application. We simulate two
 transaction types defined in TPC-C, \emph{Payment} and \emph{New order}, which together account for 88\% of the TPC-C
 workload~\cite{dbx1000}. We vary the number of warehouses from 1 to 4, and we assume that our
 transactional workload includes an equal number of transactions from both \emph{Payment} and \emph{New order}. 
TPC-H's schema consists of eight separate tables, and we use TPC-H Query 6, a long and complex workload
 that performs selection over the \emph{Lineitem} table, whose cardinality (i.e., number of rows) is 6 million. 

We evaluate the transactional and analytical throughput for \titleShort and for three baselines: (1)~SI-SS, (2)~SI-MVCC,
 (3)~MI+SW (results not shown). We find that, averaged across all warehouse counts, \titleShort has a higher transactional throughput
 (2.31X over SI-SS, 1.19X over SI-MVCC, and 1.78X over MI+SW; mean of 1.76X), and a higher analytical throughput
 (3.41X over SI-SS, 4.85X over SI-MVCC, and 2.2X over MI+SW; mean of 3.48X) over all three baselines.}

\change{We conclude that \sgii{\titleShort's ability to meet all three
HTAP properties enables better transactional and analytical performance over all three of our
state-of-the-art systems.
In Sections~\ref{sec:eval:update-propagation}, \ref{sec:eval:consistency}, and
\ref{sec:eval:analytical}, we study how each component of \titleShort
contributes to performance.}}

\subsection{Update Propagation}
\label{sec:eval:update-propagation}

\begin{figure}[h]
    \vspace{-8pt}
    \centering
        \centering
        \includegraphics[width=\linewidth]{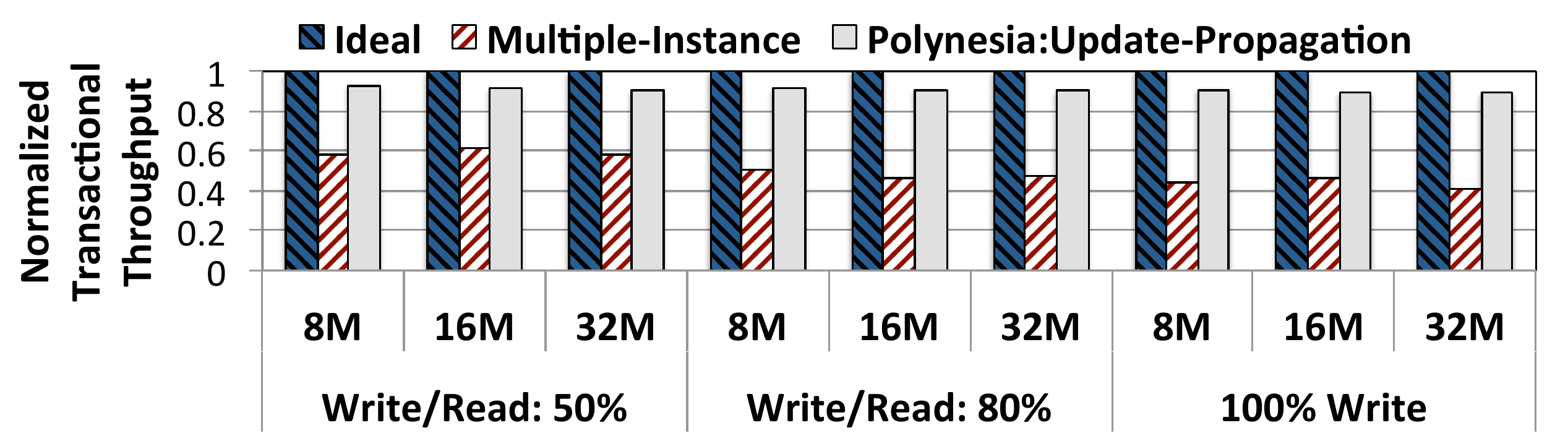}%
\vspace{-16pt}
    \caption{Effect of update propagation mechanisms on transactional throughput.}
    \label{fig:eval-update-propagation}
    \vspace{-10pt}
\end{figure}

Figure~\ref{fig:eval-update-propagation} shows
 the transactional throughput for \titleShort's update propagation
 mechanism and Multiple-Instance,
 normalized to a multiple-instance baseline with zero cost for update propagation
 (\emph{Ideal}).  
 We assume each analytical worker thread executes 
 128 queries, and vary both the number of transactional
 queries per worker thread and the
 transactional query read-to-write ratio. 
To isolate the impact of different update propagation mechanisms,
we use a zero-cost consistency mechanism, and ensure
that the level of interference remains the same for all mechanisms.

 We find that Multiple-Instance degrades transactional throughput on average
by 49.5\% compared to Ideal, as it severely suffers from resource contention and data movement cost.
27.7\% of the throughput degradation comes
from the update shipping latencies associated with data movement and with
merging updates from multiple transactional threads together.
The remaining degradation is due to the update application process,
where the major bottlenecks 
are column compression/decompression and
 dictionary reconstruction. 
 Our update propagation mechanism, on the other hand, improves throughput by 1.8X compared to
 Multiple-Instance, and comes within 9.2\% of Ideal. The improvement comes
 from (1)~significantly reducing data movement by 
 offloading update propagation process to PIM,
 (2)~freeing up CPUs from performing update propagation by using a specialized
 hardware accelerator, and (3)~tuning both hardware and software. 
In all, our mechanism reduces the latency of update
 propagation by 1.9X compared to Multiple-Instance (not shown).
We conclude that our update propagation mechanism 
provides data freshness (i.e., low update latency) while maintaining high transactional
throughput (i.e., performance isolation).

\subsection{Consistency Mechanism}
\label{sec:eval:consistency}

\begin{figure}[h]
    \vspace{-10pt}
    \centering
        \includegraphics[width=\linewidth]{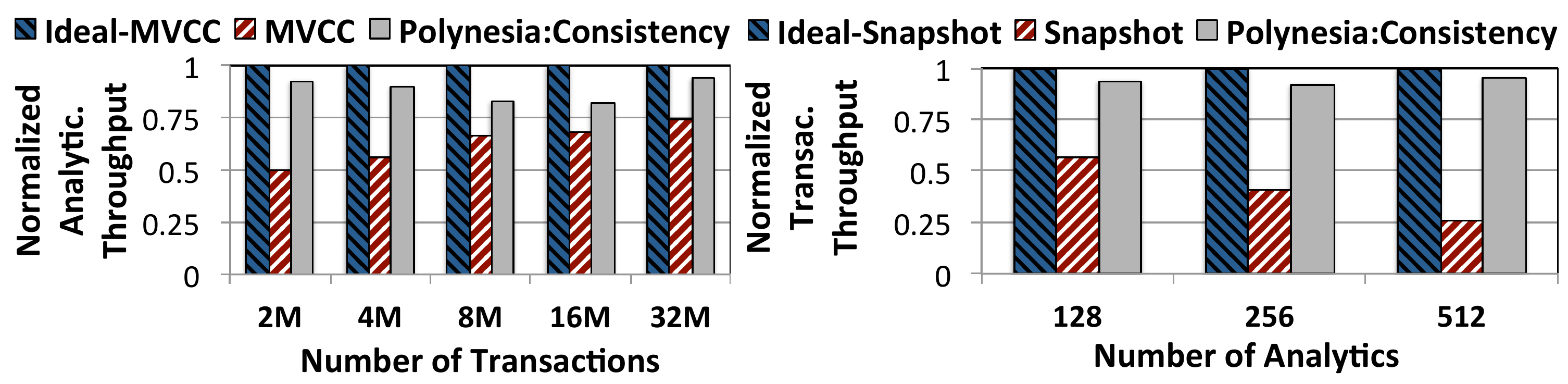}%
\vspace{-20pt}
    \caption{Effect of consistency mechanisms on analytical (left) and transactional (right) throughput.}
    \label{fig:eval-mvcc}
    \vspace{-10pt}
\end{figure}

 Figure~\ref{fig:eval-mvcc} (left) shows the analytical throughput of
 \titleShort's consistency mechanism and of Single-Instance-MVCC (\emph{MVCC}), 
 normalized to a single-instance baseline with zero cost for MVCC (\emph{Zero-Cost-MVCC})
\change{for each query count}.
 We assume each analytical worker thread executes 128 queries, and we vary
 the transactional query count per worker thread. 
 For a fair comparison, we implement our consistency 
 mechanism in a single-instance system.
\sgdel{We find that }MVCC degrades analytical throughput, on average, by 37.0\% compared to
 Ideal-MVCC, as it forces each analytical query to traverse a lengthy version
 chain and perform expensive timestamp comparisons to locate the most recent
 version. 
Our consistency mechanism, on the other hand, improves analytical throughput by 
1.4X compared to MVCC, and comes within 11.7\% of
 Ideal-MVCC, because it does
 not force analytical queries to scan lengthy version chains when accessing
 each tuple.

 Figure~\ref{fig:eval-mvcc} (right) shows
 the transactional throughput for \titleShort's consistency mechanism
and Single-Instance-Snapshot
 (\emph{Snapshot}), 
 normalized \sgii{for each workload count} to a single-instance baseline with zero-cost snapshotting (\emph{Ideal-Snapshot}).
Each worker thread performs 1M transactional queries, and we vary the number of analytical queries.
\sgdel{We find that }Snapshot reduces transactional throughput, on average, by 59\% compared
 to Ideal-Snapshot. This is because of expensive \texttt{memcpy} operations needed to create
 each snapshot, \sgii{resulting in significant resource contention}
\titleShort's mechanism improves transactional throughput by 
2.2X over Snapshot, and comes within 6.1\% of
 Ideal-Snapshot, because it \sgmod{performs snapshotting}{snapshots} at a column
 granularity and leverages PIM \sgmod{to perform}{for} fast snapshotting. 

We conclude that our consistency mechanism maintains consistency without compromising
performance isolation.

\subsection{Analytical Engine}
\label{sec:eval:analytical}

 We study the effect of each of our data placement 
 strategies from Section~\ref{sec:proposal:analytic-engine:placement}:
 Strategy~1 (\emph{Local}), Strategy~2 (\emph{Remote}), our
 \emph{Hybrid} strategy, where all use the basic scheduler heuristic.
We also study our hybrid strategy combined with our improved scheduler
 heuristic (labeled as \emph{Hybrid+sched}).
For these studies, we send all analytical queries to the same column.

\begin{figure}[h]
    \vspace{-5pt}
    \centering
        \centering
        \includegraphics[width=\linewidth]{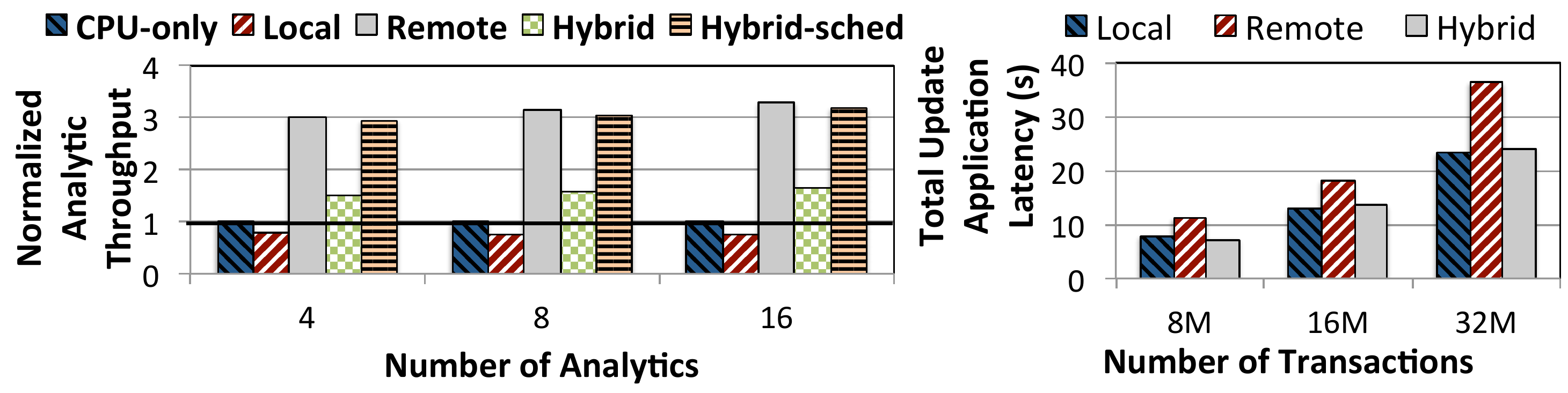}%
\vspace{-22pt}
    \caption{\sgmod{Normalized analytical throughput (left) and update application latency (right) across different data placement and task scheduling strategies.}{Effect of data placement/scheduling on throughput (left) and update application latency (right).}}
    \label{fig:eval-data-placement}
    \vspace{-10pt}
\end{figure}

Figure~\ref{fig:eval-data-placement} (left) shows the analytical throughput, \change{normalized to
the CPU-only baseline for each analytic workload count}, where one core services all queries to the same column.
We find that Local reduces throughput by 23.9\% on average over CPU-only, because in Local,
 each analytical query can only use (1)~the PIM cores in the local vault, which cannot 
 issue many memory requests concurrently, and (2)~a single vault's bandwidth.
In contrast, CPU-only leverages the out-of-order cores to issue many memory requests in parallel.
Remote improves throughput by
 4.1X/3.1X over Local/CPU-only. This is because under Remote,
 each column is partitioned across all of the vaults, allowing us to service
 each query using (1)~all of the PIM cores, and (2)~the entire internal
 bandwidth of the memory. However, Remote increases the update
 application latency, on average, by 45.8\% (Figure~\ref{fig:eval-data-placement} (right)), and thus, degrades data
 freshness. This is because of the high update application costs that we discuss in
Section~\ref{sec:proposal:analytic-engine:placement}, which Local does not incur.

We find that Hybrid addresses the shortcomings of Local, improving
 throughput by 57.2\% over CPU-only, while having a similar update application latency \change{(0.7ms)}. 
This is because the local dictionary copies eliminate most of the remote accesses.
However, the throughput under Hybrid is 49.8\% lower than Remote, because each query
 is serviced only using resources (bandwidth and computation) available
 in the local vault group.
 Hybrid-sched overcomes this \sgdel{limitation }thanks to task stealing, making idle resources in remote
vaults available for analytical queries, and comes within 3.2\% of Remote,
  while maintaining the same update application latency as Hybrid.
Note that Remote's slightly higher throughput than Hybrid-sched is because 
in Hybrid-sched, every memory access for a task stolen from another vault group is remote.

\subsection{Multiple Memory Stacks}
\label{sec:eval:multiple}

\begin{figure}[t]
    \vspace{-5pt}
    \centering
        \centering
        \includegraphics[width=\linewidth]{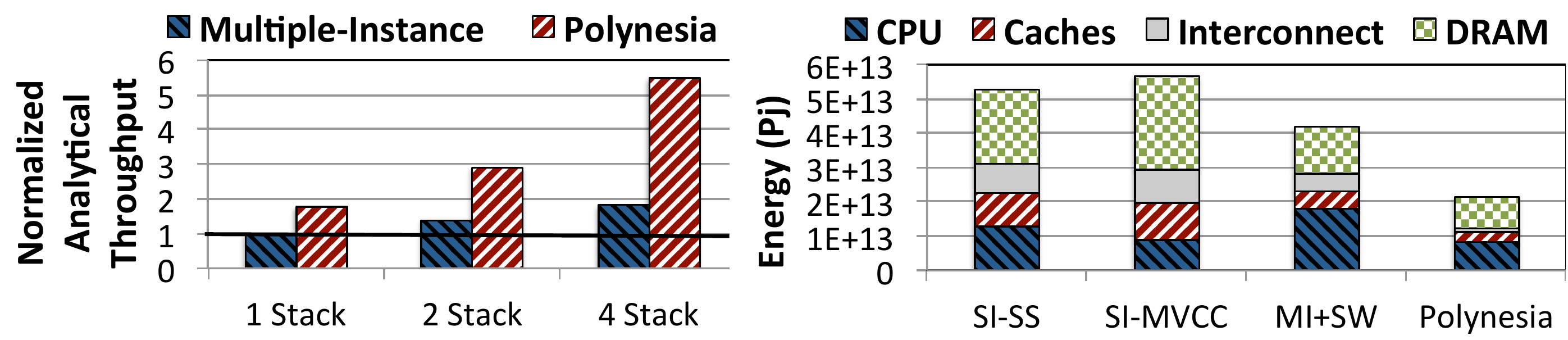}%
\vspace{-20pt}
    \caption{\change{Effect of increasing dataset size on analytical throughput (left). Total system energy (right).}}
    \label{fig:eval-energy-multistack}
    \vspace{-5pt}
\end{figure}

\change{Figure~\ref{fig:eval-energy-multistack} (left) shows how \titleShort performs as the dataset size grows.
To accommodate the larger data, we increase the number of HMC stacks, doubling the data set size as we
double the stack count.
In these studies, we use a workload with 32M transactional and 60K
 analytical queries, and analyze analytical throughput normalized to
 Multiple-Instance (MI) as a case study. We assume stacks are connected together using
 a processor-centric topology~\cite{conda}. To provide a fair comparison, we double the number
 of cores available to the analytical threads in the MI baseline as we double the number of stacks,
to compensate for the doubling of hardware resources available to \titleShort (since there
are twice as many vaults). 
We find that \titleShort significantly outperforms MI (up to 3.0X) and scales well 
as we increase the stack count. This is because, as we increase stack count, 
columns can be distributed more evenly across vault groups, which reduces the probability of multiple queries colliding
 in the same vault group. 
On the other hand, with increasing dataset size, the overheads of consistency mechanism, update propagation
 and analytical query execution are all higher for MI, which hurts its scalability.
The transactional throughput (not shown) decreases by 54.4\% at four stacks for MI, 
compared to one stack, but decreases by only 8.8\% for \titleShort.}

\subsection{Energy Analysis}
\change{Figure~\ref{fig:eval-energy-multistack} (right) shows the total system energy across different
 HTAP DBMSs. We find that MI+SW performs better than SI-MVCC and SI-SS in
 terms of energy \sgii{consumption}, but still \sgii{uses} a large amount of energy due to a large number of
 accesses to off-chip memory, large caches, and using power-hungry CPU cores. These challenges cannot
 be solved by providing high bandwidth to CPU cores. 
\sgii{\titleShort eliminates a
 significant amount of off-chip accesses, and uses custom logic and simple in-order PIM cores,
reducing its energy consumption by 48\% over MI+SW.}}

%% file: sections/related.tex

\section{Related Work}
\label{sec:related}

To our knowledge, this is the first work that 
(1)~comprehensively examines HTAP systems and their major challenges, 
(2)~proposes a hardware--software co-designed HTAP system, 
and
(3)~describes an HTAP system that meets all desired HTAP properties.
We briefly discuss related works.

\paratitle{HTAP Systems}
Several works from industry (e.g., \cite{sap-hana, oracle-dual-format, sql-htap,sap-soe, real-time-analysis-sql}) 
 and academia (e.g., \cite{hyper,peloton,hyrise,h2tap,l-store,batchdb,scyper, janus,janus-graph,sap-parallel-replication})
 propose various techniques to 
support HTAP. Many of them use a single-instance design~\cite{hyper, peloton,hyrise,h2tap,l-store,sap-hana},
while others are multiple-instance~\cite{batchdb,oracle-golden-gate,sap-soe}.
All of these proposals suffer from the drawbacks we highlight
 in Section~\ref{sec:bkgd:motiv}, and none can fully meet the desired HTAP properties.

\paratitle{Analytics Acceleration}
Other prior works focus solely on analytical 
workloads~\cite{mondrian,q100,JAFAR,widx,harp,Hash:NME}. Some of these works propose to use 
specialized on-chip accelerators~\cite{q100,widx,harp} while others propose to use PIM to 
speed up analytical operators~\cite{JAFAR,mondrian,Hash:NME}. However, none of these works study
the effect of data placement or task scheduling for the analytical workload in the context of PIM or HTAP systems.

\paratitle{PIM}
Several other recent works (e.g., \cite{google-pim, graphp, tesseract,tetris,
data-reorganization-pim, graphpim,pim-graphics, pim-enabled,tom,
Mingyu:PACT, toppim, neurocube,guo2014wondp, PICA, cairo, pattnaik.pact16})
add compute units to the logic layer of 3D-stacked memory~\cite{hbm,hmcspec2,lee2016smla, kim.cal15}.
\sgmod{While these works propose various forms of in-memory hardware, none of them}{None of these works}
are designed specifically for HTAP systems, and \sgdel{the works }are largely orthogonal.

%% file: sections/conclusion.tex

\section{Conclusion}

We propose \titleShort, a novel HTAP system that makes use of
\sgmod{multiple workload-optimized}{workload-optimized transactional and analytical} islands to enable real-time analytics
without sacrificing throughput. \sgdel{In \titleShort, transactional islands
make use of conventional transactional database engines running on
multicore CPUs, while analytical islands
make use of custom co-designed algorithms and hardware that are
placed inside memory.} Our analytical islands \sgdel{are designed to }alleviate
the data movement and workload interference costs incurred in 
state-of-the-art HTAP systems, while still ensuring that
data replicas for analytics workloads are kept up-to-date with
the most recent version of the transactional data replicas.
\sgii{\titleShort outperforms three state-of-the-art HTAP systems
(with a 1.70X/3.74X higher transactional/analytical throughput on average),
while consuming less energy (48\% lower than the best).}